\documentclass[letterpaper]{JHEP3}

\usepackage{graphicx}
\usepackage{amsmath,amsthm, amssymb}

\newcommand{\eq}{\begin{equation}}
\newcommand{\eqe}{\end{equation}}
\newcommand{\g}{\gamma}
\newcommand{\G}{\Gamma}

\newcommand{\eqa}{\begin{eqnarray}}
\newcommand{\eqae}{\end{eqnarray}}
\newcommand{\cl}{\mathcal{L}}

\title{\bf MHV Lagrangian for $N=4$ Super Yang-Mills }
\author{Haidong Feng and Yu-tin Huang\\
 {\it C.N. Yang Institute for Theoretical Physics}\\ 
{\it Stony Brook University, Stony Brook, NY 11794-3840, USA}\\
\email{hfeng@ic.sunysb.edu}\;
\email{yhuang@grad.physics.sunysb.edu}}  
\abstract{
Here we formulate two field redefinitions for N=4 Super Yang-Mills
in light cone superspace that generates only MHV vertices in the
new Lagrangian.  After careful consideration of the S-matrix equivalence theorem, we see that only the canonical transformation gives
the MHV Lagrangian that would correspond to the CSW expansion. Being in superspace, it is easier to analyse the equivalence theorem at loop level. We calculate the on shell amplitude for
4pt $(\bar{\Lambda}\bar{{\rm A}}\Lambda {\rm A})$ MHV in the new lagrangian
and show that it reproduces the previously known form. We also briefly discuss the relationship with the off-shell continuation prescription of CSW. 
}
\preprint{YITP-SB-06-50} 

\begin{document}

\numberwithin{equation}{section}


\section{Introduction} 

N=4 Super Yang-Mills theory has been an extensive field of study ever since its introduction in 1977 \cite{Sch77}. 
The large amount of symmetry has proven to be both a blessing, being a finite theory and making connections to string theory and integrability, and an obstacle. With the failure of numerous attempt to construct its off-shell formulation, in recent years the attention had turned to on-shell methods for the S-matrix of the theory, see \cite{Cachazo:2005ga, ArkaniHamed:2008gz} for review. Some of the on-shell methods developed has also been utilized in less symmetric theories. One of the ingredients is the work of Cachazo, Svrcek and Witten (CSW)\cite{Cachazo:2004kj}. From the construction of N=4 SYM tree amplitudes in terms of twistor superstring \cite{Witten:2003nn}, they propose a new perturbative approach to construct YM amplitudes based on using the on-shell form of MHV(Maximal
Helicity Violating) amplitudes as vertices. Constructed originally for YM it was also valid for N=4 SYM \cite{Georgiou:2004by}. Such an approach was also used for loop amplitudes using the cut constructible nature of N=4 SYM \cite{4}.

Various efforts has been made on providing a proof for the CSW
program. Risager \cite{Risager:2005vk} showed that the CSW program is just a
result of certain recursion relationship similar to that developed
by Britto, Cachazo and Feng \cite{Britto:2004ap}, which uses the fact that one
can use unitarity to relate one loop amplitudes to tree
amplitudes, while infrared consistency conditions relate different
tree amplitudes to satisfy a recursion relationship. However, in
the proof for the BCFW recursion relationship \cite{Britto:2005fq} one
actually uses the CSW program to prove the behavior of tree
amplitudes in certain limits. Recently, one has been able to prove that BCFW eventually leads to the CSW expansion \cite{Kiermaier:2008vz}.

Even though the relation between various on-shell methods has become clear, one would still like to see it's relationship to the action approach of QFT, since 
originally the theory was defined by its Lagrangian. Making the connection may well shed light on what properties of the Lagrangian leads to such simple structures for it's 
scatering amplitudes. Effort along this line of thought began by Gorsky and Rosly
\cite{Gorsky:2005sf} where they propose a non-local field redefinition to
transform the self-dual part of the YM action into a free action,
while the remaining vertices will transform into an infinite series
of MHV vertices. In this sense the MHV lagrangian can be viewed as a
perturbation around the self-dual sector of ordinary Yang-Mills.
This seems natural since self-dual Yang-Mills is essentially a
free theory classically. Yang-Mills lagrangian in light-cone (or
space-cone\cite{Chalmers:1998jb}) gauge is a natural framework for such a field
redefinition since the positive and negative helicity component of
the gauge field is connected by a scalar propagator. Work on the
light-cone action began by Mansfield\cite{Mansfield:2005yd} emphasizing on the
canonical nature of the field redefinition, the formulation was
also extended to massless fermions. The explicit redefinition for
Yang-Mills was worked out by Ettle and Morris \cite{Ettle:2006bw}. The
canonical condition in \cite{Mansfield:2005yd}\cite{Ettle:2006bw} ensures that using the
field redefinition complications will not arise when taking into
account of currents in computing scattering amplitude. This will
not be true for more general field redefinitions as we show in
this letter.

The progress above was mostly done in the frame work of ordinary
Yang-Mills. However, the CSW program has also achieved various
success in N=4 SYM as priorly mentioned. It is also interesting in
\cite{Ettle:2006bw} the redefinition for positive and negative helicity have
very similar form which begs for a formulation putting them on
equal footing. This formulation is present in N=4 light-cone
superspace \cite{Siegel:1981ec} where both the positive and negative helicity
gauge field sits on opposite end of the multiplet contained in a
single chiral superfield. Thus a field redefinition for one
superfield contains the redefinition for the entire multiplet,
which would be very difficult if one try the CSW program for the
component fields separately. Moreover, N=4 Self-dual YM is free at quantum level,  implying the CSW program should work better at loop level for SYM compared to YM.

In this letter we formulate such a field redefinition using the N=4
SYM light-cone Lagrangian. We proceed in two ways, first we try to
formulate a general redefinition by simply requiring the self-dual
part of the SYM lagrangian becomes free in the new Lagrangian.
Subtleties arise when using it to compute scattering amplitudes
that requires one to take into account the contribution of
currents under field redefinition. Latter, we will impose the
redefinition to be canonical. In both cases only the
redefinition of the chiral field is needed, thus giving the transformations for components in a compact manner. However, it is the second redefinition that corresponds
to CSW program, and we will see that once stripped away of the superpartners, it gives the result for YM derived in
\cite{Ettle:2006bw}. We calculate the on-shell amplitude in the new
lagrangian for 4 pt MHV amplitude and show that it matches the
simple form derived in \cite{Nair:1988bq}. In the end we briefly discuss the relation between the off-shell MHV vertices here and the on-shell form, with off-shell continuation for propagators , used in CSW.  
\section{N=4 Light-Cone Superspace} 
Without auxiliary fields susy algebra closes only up to field equations. For N=4 SYM for off-shell closure one needs infinite number of auxiliary fields which is still an area of ongoing research. However working with only on-shell degrees of freedom it is possible to manifest half of the susy. Consider N=1 SYM in d=10 
\begin{equation}
\cl=\frac{1}{g^2_{10}}tr(\frac{1}{2}F^a_{MN}F^{aMN}+\bar{\psi}\G^{M}D_M\psi)\;\;M,N=0,1,\cdot\cdot 9
\end{equation}
with transformation rules 
\begin{equation}
\delta A_M=\bar{\epsilon}\G_M\psi\;\; ;\quad \delta\psi=-\frac{1}{2}F_{MN}\G^{MN}\epsilon
\end{equation}
Consider the two subsequent susy transformation on the spinor 
\begin{eqnarray}
\nonumber(\delta_{\epsilon_2}\delta_{\epsilon_1}-\delta_{\epsilon_2}\delta_{\epsilon_1})\psi&=&-\frac{1}{2}(2\bar{\epsilon}_2\G_N D_M\psi)\G^{MN}\epsilon_1-(1\leftrightarrow2)\\
&=&(\bar{\epsilon}_2\G^P\epsilon_1)D_P\psi-\frac{1}{2}(\bar{\epsilon}_2\G^P\epsilon_1)\G_P\G^M D_M\psi
\label{closure}
\end{eqnarray} 
This closes up to the field equation $\G^M D_M\psi=0$. At this point one can still retain half of the susy on-shell\footnote{To be more precise, all susy are sill present, although half is manifested linearly and the other half non-linearly. Superspace is only useful for linear representation of susy transformation, which will be our aim here.}. In a frame where only $p^{+}$ is nonvanishing, the Dirac equation is solved if $\G^- p_-\psi=-\G^-p^+\psi=0$ where $\G^{\pm}=\frac{1}{\sqrt{2}}(\G^0\pm\G^1)$. This means that if one split the spinor $\psi$ into
\begin{equation}
\psi=-\frac{1}{2}(\G^{+}\G^{-}+\G^{-}\G^{+})\psi\equiv\psi^{+}+\psi^{-}
\end{equation}    
an on-shell spinor means that one has only $\psi^{-}$, or $\G^{+}\psi$ the ``$+$" projected spinor. Looking back at (\ref{closure}) indeed the susy algebra with $\epsilon^{+}$ closes on $\psi^{-}$. From the transformation of $A_M$ one sees that only the transverse direction transform under this reduced susy(  $\delta A_{\pm}=(\bar{\epsilon}_+\G_{\pm}\psi_-)=0$ since $\G^{-}\psi^{-}=\G^{+}\epsilon^{+}=0$). This is the basis for light-cone superfield formalism \cite{Siegel:1981ec}, where half of the susy is manifest with the on-shell degrees of freedom,  $A_{\bot}$ and $\psi^{-}$. The susy algebra one is left with is
\begin{equation}
\{Q_{\alpha+},\bar{Q}^{\beta}_{+}\}=(\g_{+})_{\alpha}\,^{\beta}p^+
\end{equation}
Preserving half of the susy means that only the SO(8) subgroup of the original Lorentz group is manifest. Dimensionally reduce to four dimensions breaks the SO(8) into SO(6)$\times$SO(2)$\sim$SU(4)$\times$ U(1). The four dimension algebra is then 
\begin{equation}
\{\bar{q}^{A},q_{B}\}=-\sqrt{2}\delta^A_B p^+
\end{equation}
where $A,B$ are SU(4) index, there are 4 complex supercharges. One can then define covariant derivatives with anti-commuting grassman variables, $\theta^A$ such that the susy generators and covariant derivatives are given by 
\begin{eqnarray}
\nonumber \bar{q}^{A}&=&-\frac{\partial}{\partial\bar{\theta}_A}-\frac{i}{\sqrt{2}}\theta^A\frac{\partial}{\partial x^ -}\;\;\;;\;\;\;\bar{d}^{A}=-\frac{\partial}{\partial\bar{\theta}_A}+\frac{i}{\sqrt{2}}\theta^A\frac{\partial}{\partial x^ -} \\ 
\nonumber q_A&=&\;\;\frac{\partial}{\partial\theta^A}+\frac{i}{\sqrt{2}}\bar{\theta}_A\frac{\partial}{\partial x^ -}\quad\;\;;\;\;d_A=\frac{\partial}{\partial\theta^A}-\frac{i}{\sqrt{2}}\bar{\theta}_A\frac{\partial}{\partial x^ -}
\end{eqnarray}
The four dimensional physical fields $\{A,\lambda^A, \phi^{AB}, \bar{\lambda}_A,\bar{A}\}$ transforms as the $\{1,4,6,\bar{4},1\}$ of SU(4). It is then natural to incorporate them in a scalar superfield, a chiral superfield  
\begin{equation}
\bar{d}^{A}\Phi=0
\end{equation} 
For N=4 SYM it's multiplet is TCP self-conjugate, therefore there
is a further constraint on the chiral fields.
\begin{equation}
\bar{\Phi}=\frac{1}{48(\partial^{+})^{2}}\epsilon^{ABCD}d_{A}d_{B}d_{C}d_{D}\Phi
\label{selfdual}
\end{equation}
which reflects the self-duality relationship of the scalar fields. Expanding in components 
\begin{eqnarray}
\Phi(x,\theta)=\frac{1}{\partial^{+}}A(y)+\frac{i}{\partial^{+}}\theta^{A}\Lambda_A(y)+i\frac{1}{\sqrt{2}}\theta^{A}\theta^{B}\bar{C}_{AB}(y)\\
\nonumber+\frac{\sqrt{2}}{3!}\theta^{A}\theta^{B}\theta^{C}\epsilon_{ABCD}\bar{\Lambda}^{D}(y)+\frac{1}{12}\theta^{A}\theta^{B}\theta^{C}\theta^{D}\epsilon_{ABCD}\partial^{+}\bar{A}(y)
\end{eqnarray}
Where
$y=(x^{+},x^{-}-\frac{i}{\sqrt{2}}\theta^{A}\bar{\theta}_{A},x,\bar{x})$ and $p^+$ appears such that each term is dimensionless. The 4 d action can then be written as 
\begin{eqnarray}
\nonumber S=tr\int d^{4}xd^{4}\theta
d^{4}\bar{\theta}\{\bar{\Phi}\frac{\partial^{+}\partial^{-}-\bar{\partial}\tilde{\partial}}{\partial^{+2}}\Phi+\frac{2}{3}gf^{abc}[\frac{1}{\partial^{+}}\bar{\Phi^{a}}\Phi^{b}\bar{\partial}\Phi^{c}+{\rm complex\;
conjugate}]\\
-\frac{g^2}{2}f^{abc}f^{ade}[\frac{1}{\partial^+}(\Phi^b\partial^+\Phi^c)\frac{1}{\partial^+}(\bar{\Phi}^d\partial^+\bar{\Phi}^e)+\frac{1}{2}\Phi^b\bar{\Phi}^c\Phi^d\bar{\Phi}^e]\}
\label{lightconeaction}
\end{eqnarray}
One can now use (\ref{selfdual}) to transform the action to depend only on the chiral superfield (chiral basis) at the expense of introducing  covariant derivatives in the interacting terms. Note however the ''self-dual" part of the action can be written in terms of only $\Phi$ quite easily.  
\section{The Field Redefinition } 
After transforming (\ref{lightconeaction}) to the chiral basis, one
arrives at a quadratic term, a three pt vertex with 4 covariant
derivatives, a three pt and four pt vertex with 8 covariant
derivatives. As shown by Chalmers and Seigel \cite{Chalmers:1996rq}, the quadratic term and the three point vertex which contains only 4 covariant derivatives describes self-dual SYM. Since self-dual SYM is free classically,  at tree level one should be able to consider the the self-dual sector to be simply a free action in the full SYM, i.e. one considers the full SYM as an perturbative expansion around the self-dual sector. Therefore the aim is to redefine the chiral field so that the self-dual sector transforms into a free action: one
then tries to find $\Phi(\chi)$ such that  
\begin{eqnarray}
S_{SD}=tr\int d^{4}xd^{4}\theta
\;\{\Phi\partial^{+}\partial^{-}\Phi-\Phi\tilde{\partial}\bar{\partial}\Phi+\frac{2}{3}\partial^{+}\Phi[\Phi,\bar{\partial}\Phi]\}\\
\nonumber=tr\int d^{4}xd^{4}\theta
\;\{\chi\partial^{+}\partial^{-}\chi-\chi\tilde{\partial}\bar{\partial}\chi\}
\label{by}
\end{eqnarray}
Note that if the field redefinition does not contain covariant derivatives, the remaining interaction terms will becomes MHV vertices, the
infinite series generated by the field redefinition from the
remaining 3 and 4 pt vertex will all have 8 covariant derivatives.
This result is implied by the known MHV amplitude \cite{Nair:1988bq}
\begin{equation*}
A(...j^{-}.....i^{-}...)_{tree}=\frac{\delta^{8}(\sum_{i=1}^{n}\lambda_{i}\theta_{i}^{A})}{\Pi^{n}_{i=1}<ii+1>}
\end{equation*}
where
\begin{equation}
\delta^{8}(\sum_{i=1}^{n}\lambda_{i}\theta_{i}^{A})=\frac{1}{2}\prod_{A=1}^{4}(\sum_{i=1}^{n}\lambda_{i}^{\alpha}\theta_{i}^{A})(\sum_{i=1}^{n}\lambda_{i\alpha}\theta_{i}^{A})
\end{equation}
The amplitude contains various combination of 8 $\theta$s and thus
imply 8 covariant derivatives to extract the amplitude.

In the Yang-Mills MHV lagrangian \cite{Mansfield:2005yd}\cite{Ettle:2006bw}, the positive
helicity gauge field $A$ transforms into a function of only the new
positive helicity field $B$, while the negative helicity $\bar{A}$
transform linearly with respect to $\bar{B}$,
$\bar{A}(\bar{B},B)$. One can see this result by noting that in order to preserve the equal time commutation relationship, 
\begin{equation}
[\partial^+\bar{A} ,A]=[\partial^+\bar{B} ,B]
\label{constraint}
\end{equation} 
that is, the field redefinition is canonical. This implies
$\partial^{+}\bar{A}=\partial^{+}\bar{B}\frac{\delta B}{\delta
A}$, therefore $\bar{A}$ transform into one $\bar{B}$ and multiple
B fields. This result for the gauge fields becomes natural in the
N=4 framework since now the chiral field $\Phi$ is redefined in
terms of series of new chiral field $\chi$. The positive helicity gauge
field A which can be defined in the superfield as
$\frac{A}{\partial^{+}}=\Phi|_{\theta=0}=\Phi(\chi|_{\theta=0})$ resulting in a function
that depends only on B. For the negative helicity
$\partial^{+}\bar{A}=D^{4}\Phi|_{\theta=0}=.....\chi (D^{4}\chi)\chi|_{\theta=0}...$,
dropping contributions from the super partners we see that
$\bar{A}(\bar{B},B)$ depends on $\bar{B}$ linearly.

Another advantage of working with superfields is that as long as
the field redefinition does not contain covariant derivatives, the
super determinant arising from the field redefinition will always
be unity due to cancellation between bosonic and fermionic
contributions. Therefore there will be no jacobian factor arising.

The requirement that the field redefinition must be canonical is necessary for the equivalence between MHV lagrangian and the original lagrangian in the framework of the LSZ reduction formula for scattering amplitudes. Indeed we will illustrate this fact by solving the field redefinition for (\ref{by}) disregarding the canonical constraint. We will show that this gives a solution that by itself does not give the correct form of MHV amplitude on-shell, one needs to incorporate the change induce on the external currents. After imposing the canonical constraint we derive the correct on-shell result. 

\subsection{Field redefinition I $\Phi(\chi)$} 
We proceed by expanding $\Phi$ in terms of $\chi$. Since the light-cone action in the component language corresponds to choosing a light-cone gauge, the redefinition should be performed on the equal light-cone time surface to preserve the gauge condition. We thus Fourier transform the remaining three coordinate into momentum space, leaving the time direction alone understanding that all fields are defined on the same time surface.
\begin{equation}
\Phi(\vec{p}_{1})=\chi(1)+\sum_{n=2}^{\infty}\int_{\vec{p}_{2}\vec{p}_{3}..\vec{p}_{n+1}}C(\vec{p}_{2},\vec{p}_{3}\cdot\cdot\cdot\vec{p}_{n+1})\chi(2)\chi(3)..\chi(n+1)\delta(\vec{p}_{1}+\sum_{i=2}^{n+1}\vec{p}_{i})
\label{super}
\end{equation}
Here we follow the simplify notation in \cite{Ettle:2006bw}, the light-cone
momentums are labelled $p=\{p^{-},p^{+},\tilde{p},\bar{p}\}$, the later
spatial momentums are collected as a three vector $\vec{p}$ and introduce abbreviation for the momentum carried by the
fields, $\chi(i)=\chi(-\vec{p}_{i})$. Plugging into (\ref{by}) the coefficient in front of the first term is determined by
equating terms quadratic in $\chi$ on the left hand side with the
right. Similarly for cubic terms we have :
\begin{eqnarray}
\nonumber\delta(\vec{p}_{1}+\vec{p}_{2}+\vec{p}_{3})tr\int d^4\theta\int_{\vec{p}_{2}\vec{p}_{3}\vec{p}_{1}}[-2C(\vec{p}_{2},\vec{p}_{3})P_{2,3}^{2}+\frac{2}{3}(p_{3}^{+}\bar{p}_{2}-p_{2}^{+}\bar{p}_{3})]\chi(1)\chi(2)\chi(3)=0\\
\end{eqnarray}
Thus we have
\begin{equation}
C(\vec{p}_{2},\vec{p}_{3})=-\frac{1\{23\}}{3P_{2,3}^{2}}
\label{solution}
\end{equation}
 Where $P_{i..j}^{2}=(p_{i}+....p_{j})^{2}$,
 $\{i,j\}=p_{i}^{+}\bar{p}_{j}-p_{j}^{+}\bar{p}_{i}$, and for later $(i,j)=p_{i}^{+}\tilde{p}_{j}-p_{j}^{+}\tilde{p}_{i}$.

For four field terms :
\begin{eqnarray}
\nonumber&&\delta(\Sigma_{i=2}^{5}\vec{p}_{i})tr\int d^4\theta\int_{\vec{p}_{2}\cdot\cdot\vec{p}_{5}}[-C(\vec{p}_{2},\vec{p}_{3})C(\vec{p}_{4},\vec{p}_{5})P_{2,3}^{2}-2C(\vec{p}_{2},\vec{p}_{3},\vec{p}_{4})P_{2,3,4}^{2}\\
\nonumber&&-\frac{2}{3}C(\vec{p}_{2},\vec{p}_{3})\{4,5\}-\frac{2}{3}C(\vec{p}_{3},\vec{p}_{4})\{(3,4),5\}-\frac{2}{3}C(\vec{p}_{4},\vec{p}_{5})\{3,(4,5)\}]\chi(2)\chi(3)\chi(4)\chi(5)=0\\
\end{eqnarray}
Using our solution for
$C(\vec{p}_{2},\vec{p}_{3})$ from (\ref{solution}), cyclic identity within trace and
relabelling the momentums for the last three terms we have:
\begin{equation}
C(\vec{p}_{2},\vec{p}_{3},\vec{p}_{4})=\frac{5}{18}\frac{\{2,3\}\{4,5\}}{P_{2,3,4}^{2}P_{2,3}^{2}}
\label{3ptsolution}
\end{equation}
One can again use this result to obtain higher terms iteratively.
The field redefinition does not contain covariant derivatives,
thus guarantees the remaining vertex after field redefinition will
be only of MHV vertex. However if we directly use the new vertices
to calculate on-shell amplitude we find that it will differ from
the original amplitude computed using the old action. In the next
subsection we use YM to illustrate the discrepancy and it's
remedy.
\subsection{Field redefinition I for YM}
One can easily follow the above procedure to solve YM field
redefinition\footnote{This redefinition was also investigated in \cite{Brandhuber:2006bf}.}. Again we have :
\begin{eqnarray}
tr\int d^{4}x \quad
\bar{A}\partial^{+}\partial^{-}A-\bar{A}\bar{\partial}\tilde{\partial}
A-\frac{\bar{\partial}}{\partial^+}A[A,\partial^+\bar{A}]\\
\nonumber=tr\int d^{4}x \quad
\bar{B}\partial^{+}\partial^{-}B-\bar{B}\bar{\partial}\tilde{\partial}
B
\end{eqnarray}
We can choose to leave $\bar{A}$ alone, $\bar{A}=\bar{B}$. Following steps similar to the above, for
the next to linear term one have:
\begin{equation}
A(1)=B(1)+\int_{\vec{p}_{2}\vec{p}_{3}}C(\vec{p}_{2},\vec{p}_{3})B(2)B(3)\delta(\vec{p}_{1}+\vec{p}_{2}+\vec{p}_{3}).....
\end{equation}
With
\begin{equation}
C(\vec{p}_{2},\vec{p}_{3})=\frac{ip_{1}^{+}\{2,3\}}{p_{2}^{+}p_{3}^{+}P_{2,3}^{2}}
\label{time}
\end{equation}
One can then use this result to compute a four point MHV amplitude. With the momentum being on shell now one has 
\begin{equation}
C(\vec{p}_{2},\vec{p}_{3})=\frac{ip_{1}^{+}}{(2,3)}
\label{answer}
\end{equation}
To see that this does not give the correct result, note that (\ref{answer}) is exactly the
required redefinition, $\Upsilon(123)$, for $A$ field derived \cite{Ettle:2006bw}. However, in
\cite {Ettle:2006bw} there is also a field redefinition for $\bar{A}$ while
in our approach we left it alone, thus it is obvious that our
redefinition will not give the correct on-shell MHV amplitude. The
difference between our approach and \cite{Ettle:2006bw} is the lacking of
canonical constraint of the field redefinition. One might guess
the discrepancy comes from the jacobian factor in the measure
generated by our redefinition (which will be present for YM). However these only
contribute at loop level. It is peculiar that field redefinition
in the lagrangian formulism should be submitted to constraints in
the canonical formulism. Direct comparison for the four pt MHV (-
-++) we see that we reproduce the last two terms in eq.(3.13)
\cite{Ettle:2006bw} while the first two terms are missing, the two terms
coming from the result of redefining the the $\bar{A}$ field.

The resolution to the missing terms comes from new contribution
arising from the currents. In a beautiful discussion of field
redefinitions in lagrangian formulism \cite{15}, it was pointed
out that since scattering amplitudes are really computed in the
lagrangian formulism with currents, one should also take into
account the effect of the field redefinition for the currents. In
the LSZ reduction formula for amplitude, one connects the source
to the Feynman diagrams being computed through propagators and
then amputate the propagator by multiplying $p^{2}$ and taking it
on-shell. For YM the currents are $J\bar{A}$ and $\bar{J}A$ where
$J$ carries the $A$ external field and $\bar{J}$ carries the $\bar{A}$
field, as can be seen by connecting them to $\langle
A\bar{A}\rangle$ propagator. When performing a field redefinition the coupling of the current with the new fields now takes a very different form
\begin{equation}
\bar{J}A(B)\rightarrow\bar{J}B+C_2\bar{J}BB+\cdot\cdot
\end{equation}
due to these higher order terms, the currents themselves behave as interaction terms. In \cite{Ettle:2006bw} these higher order contribution vanish after multiplying $p^{2}$ and taking it on-shell in the LSZ procedure. In our approach these higher terms will not vanish because of the
$\frac{1}{p^{2}}$ always sitting in front of each field
redefinition coefficient as in (\ref{solution})(\ref{3ptsolution}).
Remember the scattering amplitudes are always computed by taking
$\frac{\delta}{\delta J}$ (or  $\frac{\delta}{\delta \bar{J}}$ )of
the path integral and multiplying each $J$ (or $\bar{J}$) by $p^{2}$
and external wave function, taking everything on-shell in the end.
The non-vanishing of the additional terms means we have new
contributions to the amplitude.

Adding the contribution of these terms we shall see that one gets the correct amplitude. Consider the 4pt MHV(- -++) or
($\bar{J}\bar{J}JJ$) amplitude. Now there are four new terms
present, two for two different ways of connecting the $\bar{J}BB$
term to the original three pt.vertex, and there is two three point
vertex available. A typical graph would be that shown in fig.1,
\begin{figure} 
\centering 
\includegraphics[scale=0.8]{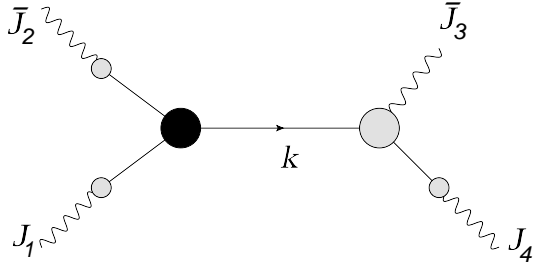}
\caption{ In this figure we show how the field redefinition may contribute to tree graphs from the modification of coupling to the source current. The solid circle indicate the $(--+)$ vertex while the empty
circle indicates contraction with the currents. Due to new terms in coupling, the $C\bar{J}_3BB$ term, one can actually construct contribution to the $(--++)$ amplitude by using this term, denoted by the larger empty circle, as a vertex. } 
\end{figure}

Consider the three pt vertex
$\frac{-i\tilde{p}_{2}}{p_{2}^{+}}p_{1}^{+}\bar{B}(k)\bar{B}(2)B(1)$
in the original lagrangian. The $\bar{B}(k)$ leg is now connected
to the $\bar{J}BB$ vertex, thus contributing a
$\frac{1}{P_{1,2}^{2}}$. From the LSZ procedure there are $p^{2}$
multiplying each current. These cancel the remaining propagators
except the $\bar{J}$ for the empty circle, the $p^{2}$ of that
current cancels the $\frac{1}{p^{2}}$ in front of the field
redefinition in (\ref{time}). Putting everything together we have for.
\begin{equation}
-\frac{\tilde{p}_{2}}{p_{2}^{+}}p_{1}^{+}\delta(\vec{p}_{k}+\vec{p}_{2}+\vec{p}_{1})\times\frac{1}{P_{1,2}^{2}}\times\frac{p_{3}^{+}\{-k,4\}}{-p_{k}^{+}p_{4}^{+}}\delta(\vec{p}_{3}+\vec{p}_{4}-\vec{p}_{k})
\end{equation}
Using the delta function and putting all external momentum
on-shell we arrive at
\begin{equation}
- \frac{\tilde{p}_{2}p_{1}^{+}p_{3}^{+2}}{p_{2}^{+}(p_{3}^{+}+p_{4}^{+})(3,4)}
\end{equation}
One can proceed the same way to generate other terms by connecting
the $\bar{B}(2)$ leg to the $\bar{J}BB$ vertex, and also doing the
same thing to the other MHV 3pt vertex
$-i\frac{\tilde{p}_{k}p_{2}^{+}}{p_{k}^{+}}\bar{B}(k)B(2)\bar{B}(3)$.
Collecting everything we reproduce the missing terms. Thus our
field redefinition does provide the same on-shell amplitude if we
take into the account of contributions coming from the currents.

\subsection{Field redefinition II (canonical redefinition)}
\label{radu}
Due to the extra terms coming from the currents, the field redefinition from the previous sections does not relate to the CSW program, since for CSW the only ingredients are the MHV vertices while above one needs current contribution. In order to avoid complication arising from the currents we impose canonical constraint as in \cite{Ettle:2006bw}, this implies the following relationship
\begin{equation}
tr\int d^{4}xd^{4}\theta
\quad\Phi(\chi)\partial^{+}\partial^{-}\Phi(\chi)
=tr\int d^{4}xd^{4}\theta
\quad\chi\partial^{+}\partial^{-}\chi
\label{giant}
\end{equation}
This is true because the canonical constraint (\ref{constraint}) implies that the new field depends on the time coordinate through the old field, there
cannot be inverse derivative of time in the coefficients that defines the redefinition. Thus our field redefinition should satisfy (\ref{giant}) and
\begin{equation}
tr\int d^{4}xd^{4}\theta
-\Phi\tilde{\partial}\bar{\partial}\Phi+\frac{2}{3}\partial^{+}\Phi[\Phi,\bar{\partial}\Phi]=tr\int d^{4}xd^{4}\theta
\;-\chi\tilde{\partial}\bar{\partial}\chi
\label{eagle}
\end{equation}
separately. To find a solution to both (\ref{giant}) and
(\ref{eagle}) one notes that the component fields are defined in the same way for both chiral superfields, we see that the A field under redefinition will not mixed with other super partners in the supersymmetric theory.
Thus we can basically read off the redefinition coefficient from
the A field redefinition derived in \cite{Ettle:2006bw}.
\begin{equation}
A(1)=B(1)+\sum_{n=2}^{\infty}-(i)^{n-1}\int_{\vec{p}_{2}\cdot\cdot\vec{p}_{n+1}}\frac{p_{1}^{+}p_{3}^{+}..p_{n}^{+}}{(23)(34).(n,n+1)}B(2)...B(n+1)\delta(\sum_{i=1}^{n}\vec{p}_{i})
\label{YM}
\end{equation}
The A field redefinition coming from the superfield redefinition in (\ref{super}) would read
\begin{equation}
\frac{A(1)}{ip_{1}^{+}}=\frac{B(1)}{ip_{1}^{+}}+\sum_{n=2}^{\infty}\int_{\vec{p}_{2}\cdot\cdot\vec{p}_{n+1}}C(2,..n+1)(i)^{n}\frac{B(2)...B(n+1)}{p_{2}^{+}..p_{n+1}^{+}}\delta(\sum_{i=1}^{n}\vec{p}_{i})
\label{SYM}
\end{equation}
Comparing (\ref{YM}) and (\ref{SYM}) implies the field redefinition for
the superfields are
\begin{equation}
\boxed{\Phi(1)=\chi(1)+\sum_{n=2}^{\infty}\int_{\vec{p}_{2}\cdot\cdot\vec{p}_{n+1}}\frac{p_{2}^{+}p_{3}^{+2}..p_{n}^{+2}p_{n+1}^{+}}{(2,3)(3,4)..(n,n+1)}\chi(2)\chi(3)..\chi(n+1)\delta(\sum_{i=1}^{n}\vec{p}_{i})
\label{SSYM}}
\end{equation}
One can check this straight forwardly by computing the redefinition for the $\bar{A}$, stripping away the superpartner contributions gives  
\begin{eqnarray}
\nonumber\bar{A}(1)&=&\bar{B}(1)+\sum_{n=2}^{\infty}\int_{\vec{p}_{2}\cdot\cdot\vec{p}_{n+1}}\sum_{s=2}^{n}(i)^{n+1}\frac{p_{s}^{+2}p_{3}^{+}p_{4}^{+}..p_{n}^{+}}{p_{1}^{+}(2,3)(3,4)..(n,n+1)}B(2)...\bar{B}(s)..B(n+1)\delta(\sum_{i=1}^{n}\vec{p}_{i})\\
\end{eqnarray}
this agrees with the result in \cite{Ettle:2006bw}. It remains to see that the solution in (\ref{SSYM}) satisfy the
constraint (\ref{giant}) and eq.(\ref{eagle}). However the fact that the pure YM
sector resulting from the super field redefinition satisfies the
constraint implies that this is indeed the correct answer. In the appendix we use this solution to prove (\ref{giant}) and eq.(\ref{eagle}) is satisfied. In the next section we use our new field redefinition to reproduce
supersymmetric MHV amplitude $\bar{\Lambda}\bar{{\rm A}}\Lambda {\rm A}$.
\subsection{ Explicit Calculation for MHV amplitude $\bar{\Lambda}\bar{{\rm A}}\Lambda {\rm A}$} 
Here we calculate the MHV
amplitude in our new lagrangian and compare to know results. For
the amplitude $\bar{\Lambda}(1)\bar{{\rm A}}(2)\Lambda(3) {\rm A}(4)$ we know
that the result is
\begin{equation}
\frac{\langle12\rangle^{2}}{\langle34\rangle\langle41\rangle}
\label{known}
\end{equation}
To transform this into momentum space we follow \cite{Ettle:2006bw}
conventions. For a massless on-shell momentum we write the spinor 
variables to be :
\begin{equation}
\lambda_{\alpha}=\left(%
\begin{array}{c}
  \frac{-\tilde{p}}{\sqrt{p^{+}}} \\
  \sqrt{p^{+}} \\
\end{array}%
\right) \;\bar{\lambda}_{\dot{\alpha}}=\left(%
\begin{array}{c}
  \frac{-\bar{p}}{\sqrt{p^{+}}} \\
  \sqrt{p^{+}} \\
\end{array}%
\right)
\end{equation}
Then we have
\begin{equation}
\langle12\rangle=\frac{(1,2)}{\sqrt{p_{1}^{+}p_{2}^{+}}}\quad
[12]=\frac{\{1,2\}}{\sqrt{p_{1}^{+}p_{2}^{+}}}
\label{map}
\end{equation}
Thus (\ref{known}) becomes
\begin{equation}
\frac{(1,2)^{2}p_{4}^{+}\sqrt{p_{1}^{+}p_{3}^{+}}}{(3,4)(4,1)p_{1}^{+}p_{2}^{+}}
\end{equation}
To compute this amplitude from our MHV Lagrangian, we use the relevant field redefinition in components, and then substitute them in the following three and four point vertex of the original Lagrangain.
\begin{equation}
-i\frac{\tilde{\partial}
\bar{{\rm A}}}{\partial^{+}}\Lambda\bar{\Lambda}+i\bar{{\rm A}}\Lambda\frac{\tilde{\partial}\bar{\Lambda}}{\partial^{+}}-i\bar{\Lambda}\frac{(\bar{{\rm A}}\Lambda)}{\partial^{+}}
{\rm A}
\label{vertex}
\end{equation}
From our field redefinition we can extract the relevant
redefinition for $\Lambda\bar{\Lambda}$
\begin{eqnarray}
\Lambda(1)\rightarrow
\int_{\vec{p}_{2}\vec{p}_{3}}i\frac{(p_{2}^{+}+p_{3}^{+})}{(2,3)}\Lambda'(2){\rm A}'(3)\delta(\vec{p}_{1}+\vec{p}_{2}+\vec{p}_{3})\\
\nonumber \bar{\Lambda}(1)\rightarrow
\int_{p_{2}p_{3}}-i\frac{p_{3}^{+}}{(2,3)}{\rm A}'(2)\bar{\Lambda}'(3)\delta(\vec{p}_{1}+\vec{p}_{2}+\vec{p}_{3})
\end{eqnarray}
Plugging into (\ref{vertex}) we have five terms. Cyclic rotate the fields
to the desired order and relabelling the momentum we arrive at
\eqa
\nonumber&&-\frac{1}{p_{2}^{+}+p_{3}^{+}}-\frac{\tilde{p}_{2}(p_{4}^{+}+p_{3}^{+})}{p_{2}^{+}(3,4)}+\frac{\tilde{p}_{1}(p_{4}^{+}+p_{3}^{+})}{p_{1}^{+}(3,4)}-\frac{p_{1}^{+}\tilde{p}_{2}}{(4,1)p_{2}^{+}}+\frac{p_{1}^{+}(\tilde{p}_{2}+\tilde{p}_{3})}{(4,1)(p_{2}^{+}+p_{3}^{+})}\\
&=&-\frac{(1,2)}{(4,1)p_2^+}-\frac{(1,2)(p_4^+ +p_3^+)}{(3,4)p_1^+p_2^+}=\frac{(1,2)^{2}p_{4}^{+}}{(3,4)(4,1)p_{1}^{+}p_{2}^{+}}
\eqae
Using the on shell external line factor in light cone for the
gauge fields is 1 and for the fermion pair is
$\sqrt{p_{1}^{+}p_{3}^{+}}$, one reproduces the MHV amplitude in
(\ref{known}).

\section{CSW-off-shell continuation}
An on-shell four momentum can be written in the bispinor form 
\begin{equation}
p_{\alpha\dot{\alpha}}=\left(\begin{array}{cc}p\bar{p}/p^+ & -\tilde{p} \\ -\bar{p} & p^+\end{array}\right)=\lambda_{\alpha}\bar{\lambda}_{\dot{\alpha}}\;\;;\;\;\lambda_{\alpha}=\left(\begin{array}{c}\frac{-\tilde{p}}{\sqrt{p^+}} \\ \sqrt{p^+}\end{array}\right),\;\bar{\lambda}_{\dot{\alpha}}=\left(\begin{array}{c}\frac{-\bar{p}}{\sqrt{p^+}} \\ \sqrt{p^+}\end{array}\right)
\label{onshell}
\end{equation}
For an off-shell momentum the relationship is modified 
\begin{equation}
p_{\alpha\dot{\alpha}}=\lambda_{\alpha}\bar{\lambda}_{\dot{\alpha}}+z\eta_{\alpha}\bar{\eta}_{\dot{\alpha}}\;\;;\;\;z=p^- -\frac{\tilde{p}\bar{p}}{p^+},\;\eta_{\alpha}=\bar{\eta}_{\dot{\alpha}}=\left(\begin{array}{c}1 \\ 0\end{array}\right)
\label{offshell}
\end{equation}  
imposing $p^2=0$ we see that $z=0$ and we are back at (\ref{onshell}). The spinors $\lambda$ and $\bar{\lambda}_{\dot{\alpha}}$ are written in terms of $p^+,\tilde{p},\bar{p}$, so that it can be directly related to amplitudes computed by the light-cone action which only contains these momentum in the interaction vertices. One can then use these spinors for the off-shell lines by keeping in mind that they relate to the momentum through (\ref{offshell}). To see this one can compute the three point MHV amplitude by looking directly at the 3 point $--+$ vertex from the light-cone action (even though these vanish by kinematic constraint, but it is sufficient to demonstrate the equivalence since the three point MHV vertex is part of the ingredient of CSW). The 3pt vertex for light-cone YM reads $i[\bar{A},p^+ A]\frac{\tilde{p}}{p^+}\bar{A}$, then the amplitude  
\begin{equation}
(1^- 2^- 3^+)=i(\frac{\tilde{p}_1}{p^+_1}p_3^+ -p_3^+\frac{\tilde{p}_2}{p^+_2})=-i\frac{p^+_3}{p^+_2 p^+_1}(1,2)=-i\frac{p^+_3}{p^+_2 p^+_1}\frac{(1,2)^3}{(2,3)(3,1)}
\end{equation}
where in the last equivalence we used $\vec{p}_1+\vec{p}_2+\vec{p}_3=0$. In our definition for the spinors, we have the identity $\langle1,2\rangle=\frac{(1,2)}{\sqrt{p^+_1 p^+_2}}$ we see that 
\begin{equation}
-i\frac{p^+_3}{p^+_2 p^+_1}\frac{(1,2)^3}{(2,3)(3,1)}=-i\frac{\langle12\rangle^3}{\langle23\rangle\langle31\rangle}
\end{equation}
Thus using this relation between the spinors and the momentum, one can relate the ``on-shell" form (in terms of $\langle ij\rangle$) to it's off-shell value (in terms of momentum).  

Now in the CSW approach the spinor for an off-shell momentum is written as $\lambda_{\alpha}=p_{\alpha\dot{\alpha}}\bar{X}^{\dot{\alpha}}$, where $\bar{X}^{\dot{\alpha}}$ is the complex conjugate spinor from an arbitrary null external line. Since in the previous analysis, one should take the identification in (\ref{onshell}) to make the connection between the MHV on-shell form and it's off-shell value, for this to work the CSW offshell continuation must be equivalent to our map, that is 
\begin{equation}
\left(\begin{array}{cc}p^- & -\tilde{p} \\ -\bar{p} & p^+\end{array}\right)\left(\begin{array}{c}\bar{X}_1 \\ \bar{X}_2\end{array}\right)=\lambda_{\alpha}=\left(\begin{array}{c}\frac{-\tilde{p}}{\sqrt{p^+}} \\ \sqrt{p^+}\end{array}\right)
\end{equation}
this leads to the requirement that  $\bar{X}^{\dot{\alpha}}=\frac{1}{\sqrt{p^+}}\left(\begin{array}{c}0 \\1\end{array}\right)$. For an arbitrary null momentum one can always find a frame such that $k_{\alpha\dot{\alpha}}=k^+\left(\begin{array}{cc}0 & 0 \\0 & 1\end{array}\right)$, this leads to $\bar{X}^{\dot{\alpha}}=\sqrt{k^+}\left(\begin{array}{cc}0  \\1\end{array}\right)$, which differs with the desired result by an overall factor $\frac{1}{\sqrt{k^+p^+}}$. This overall factor cancels in the CSW calculation since the propagator always connect two MHV graphs with one side $+$ side and the other $-$ helicity, the $+$ helicity side has a factor $(\sqrt{k^+p^+})^2$ while the negative helicity side $(\sqrt{k^+p^+})^{-2}$.

To see that one the vertices generated by the redefinition can be written in terms of the holomorphic off-shell spinors (\ref{onshell}), one needs to prove that these vertices will not depend on $\bar{p}$. This was shown in \cite{Mansfield:2005yd} to be true.   

Therefore in the MHV lagrangian, all vertices are MHV vertices and this indicates that one should be able to do perturbative calculation simply by computing Feynman graphs with only MHV vertices. Defining the map between momentum and spinor according to (\ref{onshell}), one can compute arbitrary off-shell amplitude in light-cone gauge in terms of momentum, and then map to their spinor form. Their spinor form will then take the well known holomorphic form via Nair. The difference between off-shell and on-shell is then incoded in how these spinors relate to their momentum. In a suitable basis, we see that the CSW definition for the spinor is equivalent to our on-shell off-shell map up to an overall factor that cancels in the calcualtion.   

\section{Equivalence Theorem at one-loop}
Again for this to be a proof of the CSW approach, one needs to show that the field redefinition does not introduce new terms that will survive the LSZ procedure and contribute to amplitude calculations. As discussed previously, at tree level all terms generated from the field redefinition of the coupling to source current will cancel through the LSZ procedure except the linear term. The only other possibility will be the self-energy diagram where multiplying by $p^2$ cancels the propagator that connects this diagram to other parts of the amplitude, and thus surviving.  The argument that it vanishes follows closely along the line of \cite{Ettle:2006bw}, one should be able to prove with the requirement of Lorentz invariance that all the loop integrals will be dependent only on the external momentum $p^2$ which we take to zero in the LSZ procedure. This implies that the self-energy diagrams are scaleless integrals and thus vanish.\footnote{There is of course the question of whether dimensional regularization is the correct scheme for this approach. However since in \cite{Ettle:2007qc} dimensional regularization was used to give the correct one loop amplitudes from Yang-Mills MHV Lagrangian, the analysis here should hold. However, in \cite{Brandhuber:2007vm} a different scheme was used, and it would be interesting to see if there will be any equivalence theorem violation within this scheme.  } 

We would like to compute the self-energy diagram in light-cone superspace. The Feynman rules for light-cone superspace are defined for the chiral superfield $\Phi$, thus one uses (\ref{selfdual}) to convert all the $\bar{\Phi}$ into $\Phi$. The rules have been derived in \cite{Brink:1982wv}, and here we simply use the result. \footnote{   Note that the propagators given here has already included the factor of $\bar{d}^4$ from the functional derivative $\frac{\delta\Phi(x_1,\theta_1,\bar{\theta}_1)}{\delta\Phi(x_2,\theta_2,\bar{\theta}_2)}=\frac{\bar{d}_1^4}{(4!)^2}\delta^4(x_1-x_2)\delta^4(\theta_1-\theta_2)\delta^4(\bar{\theta}_1-\bar{\theta}_2)$.}

\eqa
\nonumber\includegraphics[scale=0.8]{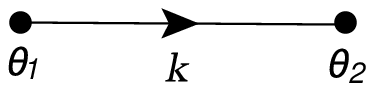}\quad&\sim&\frac{\bar{d} _1^4(k)}{k^2}\delta^8(\theta_1 -\theta_2)\\
\nonumber\includegraphics[scale=0.8]{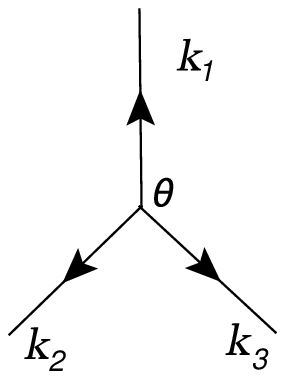}\quad&\sim&\int d^4\theta d^4\bar{\theta}\frac{d^4(p_1)}{p^{+2}_1}[\frac{1}{p_2^+}\frac{\tilde{p}_3 d^4(p_3)}{p^{+2}_3}-\frac{\tilde{p}_2 d^4(p_2)}{p^{+2}_2}\frac{1}{p_3^+}]\\
\eqae 
Here $d(k)=\frac{\partial}{\partial\theta^A}-\frac{k^+}{\sqrt{2}}\bar{\theta}_A$. The relevant graphs is now shown in fig.2.   
 
\begin{figure} 
\centering 
\includegraphics{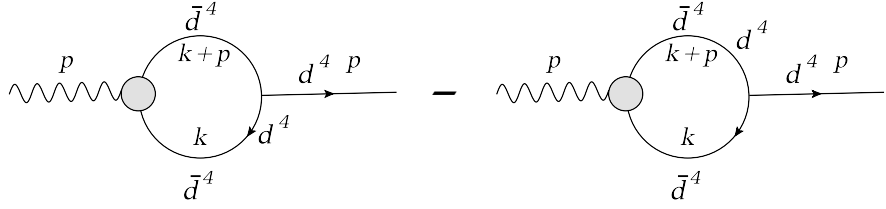}
\caption{ These are the two relevant contribution to the one-loop self-energy diagram. For simplicity we only denote the positions of $d^4$ and $\bar{d}^4$ to indicate which legs of the vertex was used for the loop contraction. } 
\end{figure}

Note that other graphs can be manipulated in to the same form by partial integrating the fermionic derivatives. Using (\ref{SSYM}) with $n=2$, the two terms give
\eqa
\nonumber&&\int d^4\theta d^4\bar{\theta}J[\frac{k^{+2}(\tilde{k}+\tilde{p})}{(k,p)k^2(k+p)^2(k^+ +p^+)}-\frac{k^+\tilde{k}}{(k,p)k^2(p+k)^2}]\Phi\\
&=&\int d^4\theta d^4\bar{\theta}J[\frac{k^{+}}{k^2(k+p)^2(k^+ +p^+)}]\Phi
\eqae
Writing in Lorentz invariant fashion we introduce a light-like reference vector $\mu$ in the $+$ direction. The result is rewritten as 
\eq
\int d^4\theta d^4\bar{\theta}J[\frac{(k\cdot\mu)}{k^2(k+p)^2(k+p)\cdot\mu}]\Phi
\eqe
Again following \cite{Ettle:2006bw}, since by rescaling $\mu\rightarrow r\mu$ the factor cancels, thus the resulting integral can only depend on $p^2$. Since we take $p^2\rightarrow0$ in LSZ reduction this means that the integral becomes a scaleless integral, and vanishes in dimensional regularization.    
\section{Discussion}
We've shown that by redefining the chiral superfield such that the
self-dual part of N=4 SYM becomes free, one generates a new
lagrangian with infinite interaction terms which are all MHV
vertex. When restricting to equal time field redefinitions the the
solution gives the suitable off-shell lagrangian that corresponds
to the CSW off-shell continuation. The redefinition is preformed
by requiring the self-dual part of the action becomes free since
the self-dual sector is essentially free classically. It does not, however, give a derivation of Nairs holomorphic form of n-point super MHV amplitude. For this purpose it is more useful to start from an action that was directly written in twistor space. Indeed such an action has been constructed in\cite{Boels:2007qn} and it's relation to CSW has been discussed.
\acknowledgments
It is a great pleasure to thank  Simone Giombi, Riccardo Ricci and
Diego Trancanelli for their discussion at the early stage of this
work, Warren Siegel for suggestion and supervision, George Sterman
and the especially James H. Ettle and Tim R. Morris for their
generous comments. This work is supported in part by the National
Science Foundation, grants PHY-0354776.
\appendix
\section{ Proof of field redefinition} 

Here we will prove that our field redefinition introduced in sec.(\ref{radu}) satisfies both (\ref{giant}) and (\ref{eagle}). We first produce the proof at leading, terms with three new fields on the LHS of both equations vanish. From this experience we will then show that the same holds for all higher order terms, namely, written in terms of new fields, terms that are more than quadratic in $\chi$ on LHS of these equations vanish. 

For (\ref{giant}) terms with three field comes from the second order term in the field redefinition, namely $\phi(1)\rightarrow C(2,3)\chi(2)\chi(3)$ with $C(2,3)=\frac{p_2^+p_3^+}{(2,3)}$, they give 
\begin{equation}
tr\int_{\vec{p}_{1}\vec{p}_{2}\vec{p}_{3}}p^-[\frac{p_{1}^{+}p_{2}^{+}p_{3}^{+}}{(1,2)}-\frac{p_{1}^{+}p_{2}^{+}p_{3}^{+}}{(2,3)}]\Phi(1)\Phi(2)\Phi(3)\delta(\Sigma_{i}\vec{p}_{i})
\end{equation}
Using momentum conservation, $(1,2)=-(3,2)=(2,3)$, these two terms indeed cancel each other. The 3 field term that is generated on the LHS for (\ref{eagle})
\begin{eqnarray}
tr\int_{\vec{p}_{1}\vec{p}_{2}\vec{p}_{3}}
-\frac{p_{2}^{+}p_{3}^{+}p_{1}\bar{p_{1}}}{(2,3)}\chi(1)\chi(2)\chi(3)+\frac{(\bar{p}_{2}p_{3}^{+}-\bar{p}_{3}p_{2}^{+})}{3}\chi(1)\chi(2)\chi(3)
\label{proof1}
\end{eqnarray}
Using cyclic identity and relabelling the momentum for the first
term we have
\eqa
\nonumber &&tr\int_{\vec{p}_{1}\vec{p}_{2}\vec{p}_{3}}-\chi(1)\chi(2)\chi(3)\frac{1}{3}[\frac{p_{2}^{+}p_{3}^{+}\tilde p_{1}\bar{p_{1}}}{(2,3)}+\frac{p_{1}^{+}p_{2}^{+}\tilde p_{3}\bar{p_{3}}}{(1,2)}+\frac{p_{3}^{+}p_{1}^{+}\tilde{p}_{2}\bar{p_{2}}}{(3,1)}]\\
\nonumber=&&tr\int_{\vec{p}_{1}\vec{p}_{2}\vec{p}_{3}}-\chi(1)\chi(2)\chi(3)[\frac{p_{2}^{+}p_{3}^{+}\tilde{p}_{2}\bar{p}_{3}+p_{2}^{+}p_{3}^{+}\tilde{p}_{3}\bar{p}_{2}-p_{2}^{+2}\tilde{p}_{3}\bar{p}_{3}-p_{3}^{+2}\tilde{p}_{2}\bar{p}_{2}}{3(2,3)}]\\
=&&tr\int_{\vec{p}_{1}\vec{p}_{2}\vec{p}_{3}}\chi(1)\chi(2)\chi(3)\frac{\{2,3\}}{3}
\eqae
where in the last two lines we used momentum conservation. This gives the same term as the second term in (\ref{proof1}) with a minus sign.

To prove that higher field terms also cancel in (\ref{giant}) for our field redefinition, note that for $n$-fields the coefficients combine into
\eqa
\nonumber&&\sum_{j=3}^{n-1}C(2,\cdot\cdot,j)p_{(j+1,n)}^+C(j+1,\cdot\cdot,n)\\
&=&-\frac{(\prod_{i=2}^{n}p_i^+)(\sum_{j=3}^{n-2}S_j)}{(2,3)(3,4)\cdot\cdot(n,n-1)}
\eqae 
where we've used the notation that $p_{(1,n)}^+\equiv\sum_{i=1}^{n}p_i^+$ and 
\eq
S_j\equiv p_{n-j}^+\cdot\cdot p_4^+p_3^+[p_{n-1}^+\cdot\cdot p_{n+3-j}^+p_{n+2-j}^+(n+1-j,n-j)+{\rm cyclic\;rotations }]
\eqe
For example for $n=7$
\eqa
\nonumber S_3&=&p_4^+p_3^+[p_6^+(5,4)+p_5^+(4,6)+p_4^+(6,5)]\\
\nonumber S_4&=&p_3^+[p_6^+p_5^+(4,3)+p_5^+p_4^+(3,6)+p_4^+p_3^+(6,5)+p_3^+p_6^+(5,4)]\\
\nonumber S_5&=&[p_6^+p_5^+p_4^+(3,2)+p_5^+p_4^+p_3^+(2,6)+p_4^+p_3^+p_2^+(6,5)+p_3^+p_2^+p_6^+(5,4)+p_2^+p_6^+p_5^+(4,3)]\\
\eqae
The important point is since these $S_j$ are cyclic sums over terms that are partially anti-symmetric, $S_j=0$. Hence we've proven that (\ref{giant}) is indeed satisfied. 

Moving on to (\ref{eagle}), we use the fact that since (\ref{giant}) is satisfied, this implies that\footnote{ Written in this form we neglect the superspace delta functions and spinor derivatives that usually arises, since we know that the chiral superfield $\Phi$ is now already written in terms of chiral superfield $\chi$.} 
\eq
\partial^+\Phi=\frac{\delta\chi}{\delta\Phi}\partial^+\chi.
\eqe
From the discussion above we see that this is indeed true. Plugging back into \ref{eagle} we have 
\eq
\frac{1}{\partial^+}[\partial^+\Phi,\bar{\partial}\Phi]=-\frac{\bar{\partial}\tilde{\partial}}{\partial^+}\Phi+\frac{\delta \Phi}{\delta\chi}\frac{\bar{\partial}\tilde{\partial}}{\partial^+}\chi
\eqe
Fourier transform into momentum space and plugging in (\ref{SSYM}) we have 
\eqa
\nonumber(-\frac{\tilde{p}_1\bar{p}_1}{p_1^+}+\sum_{i=2}^n\frac{\tilde{p}_i\bar{p}_i}{p_i^+})C(2,3,\cdot\cdot\cdot,n)&=&\frac{1}{p_1^+}\sum_{j=2}^{n}C(2,\cdot\cdot\cdot,j)C(j+1,\cdot\cdot\cdot,n)\{p_{(j+1,n)},p_{(2,j)}\}\\
&=&\frac{1}{p_1^+}\sum_{j=2}^{n}C( 2,3,\cdot\cdot\cdot,n)\frac{(j,j+1)}{p_j^+p_{j+1}^+}\{p_{(j+1,n)},p_{(2,j)}\}
\label{last}
\eqae
again $\{p_{(j+1,n)},p_{(2,j)}\}=p_{(j+1,n)}^+\bar{p}_{(2,j)}-\bar{p}_{(j+1,n)}p_{(2,j)}^+$. Since $\frac{(j,j+1)}{p_j^+p_{j+1}^+}=\frac{\tilde{p}_{j+1}}{p_{j+1}^+}-\frac{\tilde{p}_j}{p_j^+}$ the RHS becomes 
\eqa
\nonumber&&\frac{1}{p_1^+}\sum_{j=2}^{n}C( 2,3,\cdot\cdot\cdot,n)[\frac{\tilde{p}_{j+1}}{p_{j+1}^+}-\frac{\tilde{p}_j}{p_j^+}]\{p_{(j+1,n)},p_{(2,j)}\}\\
\nonumber=&&\frac{1}{p_1^+}\sum_{j=2}^{n}C( 2,3,\cdot\cdot\cdot,n)\frac{\tilde{p}_j}{p_j^+}[\{p_{(j,n)},p_{(2,j-1)}\}-\{p_{(j+1,n)},p_{(2,j)}\}]\\
=&&\frac{1}{p_1^+}\sum_{j=2}^{n}C( 2,3,\cdot\cdot\cdot,n)\frac{\tilde{p}_j}{p_j^+}\{1,j\}
\eqae
momentum conversation then gives the LHS of (\ref{last}).


\end{document}